\documentstyle[11pt,newpasp,twoside,epsf]{article}
\def\edcomment#1{\iffalse\marginpar{\raggedright\sl#1\/}\else\relax\fi}
\reversemarginpar

\begin{document}
\title{Iron Emission in {\it z} $\approx$ 6 QSOs and its Possible Implications}
\author{Michael R. Corbin}
\affil{Computer Sciences Corporation / Space Telescope Science Institute, 3700 San Martin Drive, Baltimore, MD 21218; corbin@stsci.edu}
\author{Wolfram Freudling}
\affil{Space Telescope - European Coordinating Facility, European Southern Observatory, Karl Schwarzchild Strasse 2, 85748 Garching, Germany; wfreudli@eso.org}
\author{Kirk T. Korista}
\affil{Physics Department, Western Michigan University, 1120 Everett Tower, Kalamazoo, MI 49008-5252; kirk.korista@wmich.edu}

\begin{abstract}

We have obtained low-resolution near-infrared spectra of five SDSS QSOs at
5.3 $<$ {\it z} $<$ 6.3 using the NICMOS instrument of the {\it Hubble Space Telescope}.  We find evidence of emission in the Fe II complex centered near $2500${\AA} (rest) in all five objects.  We estimate Fe II / Mg II $\lambda$2800 flux ratios comparable to those measured in QSOs at lower redshifts, which indicate metallicities near or above the solar level. We discuss the possible implications of this result assuming the iron enrichment to have been produced mainly by Type Ia supernovae.    

\end{abstract}

\section{Introduction}

The relative abundances of iron and $\alpha$-process elements in QSOs could provide an important constraint on the star-formation histories of their host galaxies (e.g. Yoshii, Tsujimoto \& Kawara 1998; Korista et al., these proceedings).  We report here the results of a Cycle 11 {\it Hubble Space Telescope} program to search for Fe II emission in five {\it z} $\approx$ 6 QSOs discovered in the Sloan Digital Sky Survey using the Near Infrared Camera and Multi-Object Spectrometer (NICMOS), and discuss the possible implications of our findings.  The results for the first three objects observed in this program are also presented in Freudling, Corbin \& Korista (2003).  

\section{Observations \& Results}
Observations of our sample objects were obtained between 2002 October 25 and 2003 June 19 using the NICMOS grism G141, with a nominal resolution of 200 pixel$^{-1}$.  This grism offers spectral coverage between approximately 1.1 $\mu$m $-$ 1.9 $\mu$m, and HST observations avoid the telluric absorption between the {\it H} and {\it K} bands.  Total integration times for each object were approximately 40 minutes, and consisted of four separate integrations with the QSOs placed on each of the NICMOS Camera 3 detector quadrants in order to minimize pixel-to-pixel and large-scale sensitivity variations across the detector. Details of the calibration and reduction procedure are given in Freudling et al. (2003). 

Our NICMOS spectra cover the portion of the rest-frame object spectra containing Al III $\lambda$1892, C III] $\lambda$1909, the Fe II $2500${\AA} emission complex, and, except for the objects at {\it z} = 6.03 and {\it z} = 6.27, Mg II $\lambda$2800. In order to estimate the Fe II / Mg II ratio, we fit our final spectra as a combination of a power-law continuum, Fe II emission using a scaled version of the composite QSO spectrum of Zheng et al. (1997) as a template, and Gaussian profiles for the Al III $\lambda$1892, C III] $\lambda$1909 and Mg II $\lambda$2800 lines.  To compare our measurements of the Fe II / Mg II ratio with those measured for objects at lower redshifts, we apply a scaling factor to account for the Fe II emission out to 3300{\AA} (rest), which our spectra do not cover (see Freudling et al. 2003).  For SDSS J1306+0356 ({\it z} = 6.03) and SDSS J1030+0524 ({\it z} = 6.27), we estimate the flux in Mg II $\lambda$2800 from that measured in C III] $\lambda$1909, using a Mg II $\lambda$2800 /  C III] $\lambda$1909 ratio of 1.36 $\pm$ 0.71, measured from a large sample of QSOs at {\it z} $\approx$ 2 (Brotherton et al. 1994).  This assumes a similar ratio in our sample objects, which is reasonable given that their spectra near Ly$\alpha$ do not show any anomalous line strengths (e.g. Fan et al. 2001; Pentericci et al. 2002), and that both C and Mg are $\alpha$-process elements.

\begin{figure}
\plotone{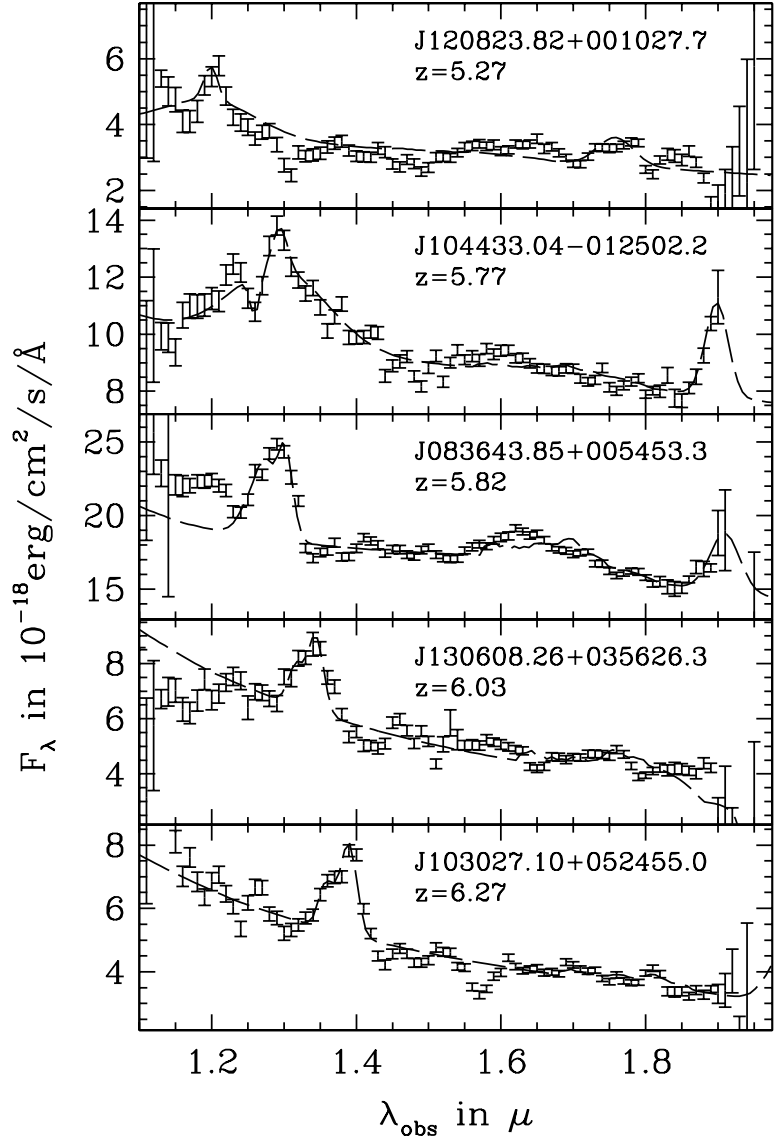}
\caption{NICMOS spectra of our sample objects. The emission line at the blue end of the spectra is the Al III $\lambda$1892, C III] $\lambda$1909 blend, and the line at the red end of the first three spectra is Mg II $\lambda$2800.  The dashed lines represent the best fits to the spectra as described in the text.  Note the strength of the Fe II emission in SDSS J0836+0054.}
\end{figure}

\begin{table}
\caption{Fe II / Mg II Flux Ratios}
\begin{tabular}{cccc}
\tableline
\tableline
name & {\it z} & {\it F}(Fe II)/{\it F}(Mg II) & $\sigma$[{\it F}(Fe II)/{\it F}(Mg II)] \\
\tableline
SDSS J0836+0054 & 5.82 & 10.3 & 2.2 \\
SDSS J1030+0524 & 6.27 & 3.2 & 1.2 \\
SDSS J1044$-$0125 & 5.77 & 4.6 & 1.9 \\
SDSS J1208+0010 & 5.27 & 4.5 & 2.9 \\
SDSS J1306+0356 & 6.03 & 5.0 & 2.2 \\
\tableline
\end{tabular}
\end{table}
 
Our spectra and best fits to them are shown in Figure 1.  Table 1 presents the corresponding scaled values of the Fe II / Mg II flux ratios, their uncertainties, and the redshifts we measure from the best fits to the emission lines. Our measurement of the ratio in SDSS J1208+0010 is very uncertain. The spectrum of Iwamuro et al. (2002) shows the Mg II profile to be very narrow, and it is unresolved in our spectrum, making it difficult to separate from the Fe II emission.  We note that while the Fe II $2500${\AA} complex is weak in both SDSS J1306+0356 and SDSS J1030+0524, our best fits still yield some flux in it.  

Our Fe II / Mg II flux ratios fall within the range measured for objects at lower redshifts (e.g. Iwamuro et al. 2002; Dietrich et al. 2002).  Barth et al. (2003, and these proceedings) also find strong Fe II emission near Mg II $\lambda$2800 in the {\it z} = 6.4 QSO SDSS J1148+5251, and estimate a total Fe II / Mg II ratio within the same range.  There is thus no evidence of a change in this ratio with redshift. Although the Fe II emission of AGN has proven difficult to model (see Baldwin et al. these proceedings), these ratios are consistent with Fe and Mg abundances near or above the solar level (Wills, Netzer \& Wills 1985).  
\section{Discussion}

If the iron detected in {\it z} $\approx$ 6 QSOs were produced mainly by SNe Ia, the evidence that it has reached solar and super-solar abundance levels has several important implications.  First, the galaxy mass / metallicity relation indicates that the host galaxies of the QSOs are very massive, possibly the progenitors of giant ellipticals (see Hamann \& Ferland 1999 and Fan et al., these proceedings). Such massive objects thus formed within $\sim$ 1 Gyr of the Big Bang.  Second, recent calculations of SNe Ia formation and ISM enrichment timescales (Friaca \& Terlevich 1998) indicate that solar Fe levels can be reached within 0.5 to 0.8 Gyr after the formation of the progenitor stars. Under currently favored cosmologies ({\it h} $\approx$ 0.7, $\Omega_{\Lambda}$  $\approx$ 0.73, $\Omega_{M}$ $\approx$ 0.27) our results imply that the progenitor stars formed at {\it z} $\approx$ 20 $\pm$ 10 (see Freudling et al. 2003). This is consistent with the estimate of the first reionization epoch made from the polarization of the cosmic microwave background (Bennett et al. 2003).  Finally, indirect and direct (Haiman \& Cen 2002) constraints on the ages of z $\approx$ 6 QSOs indicate that they reached their present luminosities near their observed redshifts.  The evidence that the stars of their host galaxies formed at higher redshifts thus supports models of an evolutionary connection between nuclear star formation and the AGN phenomenon (e.g. Norman \& Scoville 1988; Kauffmann \& Haenelt 2000; Kawakatu \& Umemura 2003).

\end{document}